\documentclass[sigconf]{acmart}
\usepackage[frozencache,cachedir=.]{minted}
\usepackage{bm}
\usepackage{lipsum}
\usepackage{algorithm}
\usepackage{algpseudocode}
\usepackage{xcolor}
\usepackage{multicol}
\usepackage{multirow}
\usepackage{newtxmath}
\usepackage{caption}
\usepackage{graphicx}
\usepackage{subfigure}
\usepackage{amsmath}
\usepackage{balance}
\usepackage{marvosym}
\newcommand{\x}{{\bf x}}

\newcommand{\z}{{z}}

\AtBeginDocument{%
  \providecommand\BibTeX{{%
    \normalfont B\kern-0.5em{\scshape i\kern-0.25em b}\kern-0.8em\TeX}}}

\copyrightyear{2023} 
\acmYear{2023} 
\setcopyright{acmlicensed}\acmConference[KDD '23]{Proceedings of the 29th ACM SIGKDD Conference on Knowledge Discovery and Data Mining}{August 6--10, 2023}{Long Beach, CA, USA}
\acmBooktitle{Proceedings of the 29th ACM SIGKDD Conference on Knowledge Discovery and Data Mining (KDD '23), August 6--10, 2023, Long Beach, CA, USA}
\acmPrice{15.00}
\acmDOI{10.1145/3580305.3599851}
\acmISBN{979-8-4007-0103-0/23/08}

\begin{document}

\title{Joint Optimization of Ranking and Calibration with Contextualized Hybrid Model}


\author{Xiang-Rong Sheng}
\email{xiangrong.sxr@alibaba-inc.com}
\affiliation{%
  \institution{Alibaba Group}
    \city{Beijing}
    \country{China}
}

\author{Jingyue Gao}
\email{jingyue.gjy@alibaba-inc.com}
\affiliation{%
  \institution{Alibaba Group}
    \city{Beijing}
    \country{China}
}

\author{Yueyao Cheng}
\email{chengyueyao.cyy@alibaba-inc.com}
\affiliation{%
  \institution{Alibaba Group}
    \city{Beijing}
    \country{China}
}

\author{Siran Yang}
\email{siran.ysr@alibaba-inc.com}
\affiliation{%
  \institution{Alibaba Group}
    \city{Beijing}
    \country{China}
}

\author{Shuguang Han}\authornote{Shuguang Han is the corresponding author.}
\email{shuguang.sh@alibaba-inc.com}
\affiliation{%
  \institution{Alibaba Group}
    \city{Beijing}
    \country{China}
}

\author{Hongbo Deng}
\email{dhb167148@alibaba-inc.com}
\affiliation{%
  \institution{Alibaba Group}
    \city{Beijing}
    \country{China}
}

\author{Yuning Jiang}
\email{mengzhu.jyn@alibaba-inc.com}
\affiliation{%
  \institution{Alibaba Group}
    \city{Beijing}
    \country{China}
}

\author{Jian Xu}
\email{xiyu.xj@alibaba-inc.com}
\affiliation{%
  \institution{Alibaba Group}
    \city{Beijing}
    \country{China}
}

\author{Bo Zheng}
\email{bozheng@alibaba-inc.com}
\affiliation{%
  \institution{Alibaba Group}
    \city{Beijing}
    \country{China}
}

\renewcommand{\shortauthors}{Sheng et al.}

\begin{abstract}
Despite the development of ranking optimization techniques, pointwise loss remains the dominating approach for click-through rate prediction. It can be attributed to the \textbf{calibration ability} of the pointwise loss since the prediction can be viewed as the click probability. In practice, a CTR prediction model is also commonly assessed with the \textbf{ranking ability}. To optimize the ranking ability, ranking loss (e.g., pairwise or listwise loss) can be adopted as they usually achieve better rankings than pointwise loss. Previous studies have experimented with a direct combination of the two losses to obtain the benefit from both losses and observed an improved performance. However, previous studies break the meaning of output logit as the click-through rate, which may lead to sub-optimal solutions. To address this issue, we propose an approach that can \textbf{J}ointly optimize the \textbf{R}anking and \textbf{C}alibration abilities (\textbf{JRC} for short). JRC improves the ranking ability by contrasting the logit value for the sample with different labels and constrains the predicted probability to be a function of the logit subtraction. We further show that JRC consolidates the interpretation of logits, where the logits model the joint distribution. With such an interpretation, we prove that JRC approximately optimizes the contextualized hybrid discriminative-generative objective. Experiments on public and industrial datasets and online A/B testing show that our approach improves both ranking and calibration abilities. Since May 2022, JRC has been deployed on the display advertising platform of Alibaba and has obtained significant performance improvements.
\end{abstract}


\maketitle

\section{Introduction}
Click-through rate (CTR) prediction has played a vital role in many industrial applications, such as recommender systems and online advertising. A CTR prediction model is commonly evaluated from two perspectives: whether the prediction aligns with the actual click-through rate (i.e., calibration ability)~\cite{GuoPSW2017OnCalibration,MindererDRHZHTL2021RevisitCalibration}, and whether the prediction leads to a correct ranking (i.e., ranking ability)~\cite{GuoPSW2017OnCalibration,li2015click}.

Despite recent progress in ranking optimization techniques, the pointwise model~\cite{HePJXLXSAHBC14PracticalLessonsFB,graepel2010web,cheng2016wide,zhou2018din,zhang2022keep} remains the dominating approach for CTR prediction. The wide use of the pointwise model is mainly attributed to its calibration ability since the prediction can be treated as the click probability~\cite{HastieFT2001ESL}.
\begin{figure*}[!h]
	\centering
	\includegraphics[width=\linewidth]{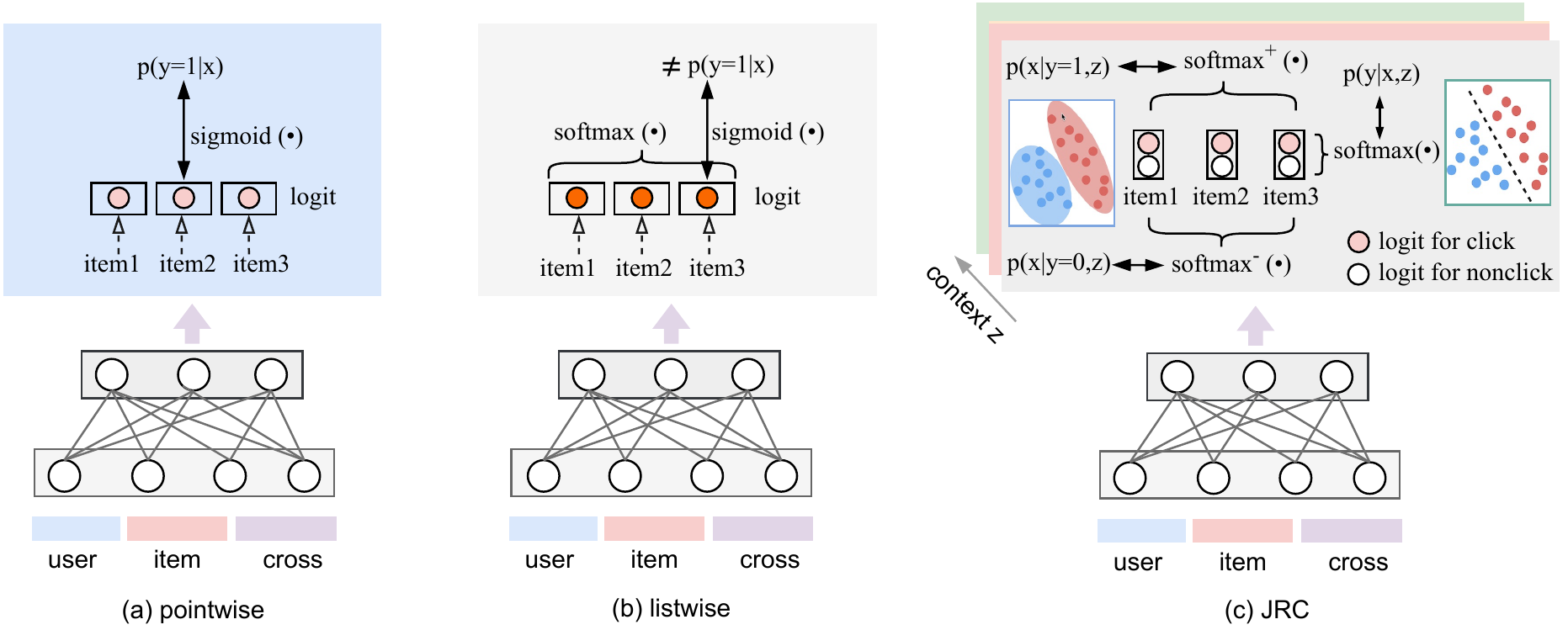}
	\caption{A comparison among the three approaches: (a) pointwise model, (b) listwise model, and (c) our proposed JRC model.}
	\label{fig:hybrid}
\end{figure*}
However, in the pointwise model, direct optimization of the cross-entropy loss may fail to improve the ranking ability. Such a loss function does not consider the discriminative ability among items, which, on the contrary, is the primary measurement for the ranking ability~\cite{liu2011learning}. In addition, the identical and independently distributed (i.i.d.) assumption in the pointwise model may deviate from real-world applications. For example, users typically see a list of items and make click decisions. Therefore, researchers have proposed various pairwise and listwise approaches to directly optimize ranking metrics such as AUC (Area Under the Curve)~\cite{Fawcett2006ROC}, nDCG (normalized discounted cumulative gain)~\cite{jarvelin2002cumulated}, MAP (Mean Average Precision)~\cite{baeza1999modern}. They improve the ranking-related metrics to a large extent but introduce new issues.

With a pairwise or listwise approach, the predictions will no longer correspond to click probability~\cite{liu2011learning,bruch2020stochastic}. In general, logits\footnote{In this research, the logit refers to the model output \emph{before} the last activation layer (e.g., sigmoid or softmax function).} in the pointwise and pairwise/listwise methods are defined based on different underlying assumptions. As illustrated by Figure \ref{fig:hybrid}(a), the logit in a pointwise model can be easily converted (after sigmoid transformation) to the click probability,
whereas the logit in a listwise model, e.g., ListNet~\cite{CaoQLTL2007ListNet}, only measures the degree of relevance since the softmax function is conducted over all of the items within the same list, lacking a direct connection to the click probability. 
To preserve the benefits from both pointwise and pairwise/listwise approaches, an intuitive way, is to combine the two loss functions~\cite{sculley2010combined,li2015click}. Despite being effective, the meaning of the integrated logit becomes unclear as it is adopted to define two different probabilities, i.e., the probability of click and the probability of being ranked on the top. 

The abovementioned issues call for the redesign of logit to optimize ranking and calibration within the same framework.
To this end, we propose a \textbf{J}oint optimization approach that maximizes the \textbf{R}anking and \textbf{C}alibration abilities (\textbf{JRC} for short) at the same time. To avoid carrying different meanings using one single logit, we extend the one-dimensional logit to two-dimensional logits, i.e., adding another degree of freedom to the model output. Let $f_\theta$ denote the model parameterized by $\theta$, the output of this model is a vector of two values $f_{\theta}(\x)[0]$ and $f_{\theta}(\x)[1]$.  $f_{\theta}(\x)[0]$ and $f_{\theta}(\x)[1]$ represent the logits corresponding to the non-click state and click state, respectively. Hereafter, we refer them as to the click-logit and the nonclick-logit.

With these two logits, we may simply adopt a multi-task modeling approach that uses one for ranking optimization and the other for calibration optimization. However, in this way, such two logits are connected indirectly, making it hard to fuse them for the final prediction.
On contrary, as shown in Equation~\ref{eq:logit-subtraction}, JRC computes the predicted click probability based on the subtraction $f_{\theta}(\x)[1] - f_{\theta}(\x)[0]$, in which both of the two logits contributed to the final prediction. Here, the predicted click probability is obtained by applying the sigmoid function on $f_{\theta}(\x)[1] - f_{\theta}(\x)[0]$, which is equivalent to applying the softmax function over the two logits.
\begin{equation}
\begin{aligned}
  \hat{p}(y=1|\x) &=
     \frac{1}{1+\exp( -(f_{\theta}(\x)[1] - f_{\theta}(\x)[0]))} \label{eq:logit-subtraction}
\end{aligned}
\end{equation}

As for the optimization objective, we firstly introduce a pointwise loss, using the two logits, to preserve the calibration ability:
\begin{equation}
\begin{aligned}
  \ell_{\text{calib}} &= -\sum_{{\x,y}} \mathrm{log} ~\hat{p}(y|\x) \\
  &= -\sum_{{\x,y}} \log \frac{\exp(f_{\theta}(\x)[y])}
       {\exp(f_{\theta}(\x)[0]) + \exp(f_{\theta}(\x)[1])}.~\label{eq:calib-loss}
\end{aligned}
\end{equation}

To further improve the ranking ability, we adopt a listwise-like loss. For each positive sample $(\x, y=1)$, the listwise-like loss contrasts its click-logit to the click-logits of all other samples within the same context $z$. 
For each negative sample $(\x, y=0)$, the listwise-like loss encourages its nonclick-logit to be larger than the nonclick-logits of all other samples.
Such a process can be illustrated by Equation~\ref{eq:listwise-ranking},
\begin{equation}
 \begin{aligned}
  \ell_{\text{rank}} = 
    -\sum_{\x, y, z} \log\frac{\exp(f_{\theta}(\x)[y])}{\sum_{\x_i\in X_{\z}}\exp (f_{\theta}(\x_i)[y])},
 \end{aligned}
 \label{eq:listwise-ranking}
\end{equation}
where $\z$ indicates the current context, e.g., the current session~\cite{li2015click}. $X_{\z}$ denote the set of samples that share the same context $\z$.
The final objective of JRC is defined as
\begin{equation}
 \begin{aligned}
  \ell_{\text{final}} = \alpha \ell_{\text{calib}} + (1-\alpha) \ell_{\text{rank}},
 \end{aligned}
 \label{eq:final-obj}
\end{equation}
in which $\alpha$ is the hyper-parameter that balances the importance of different loss functions.

By introducing an additional degree of freedom, the above formulation alleviates the conflict of optimizing ranking and calibration with only one single logit. 
We further show that JRC consolidates the interpretation of logits for both pointwise loss and listwise-like loss. Particularly, JRC can be seen as an energy-based model (EBM)~\cite{lecun2006tutorial}, in which logits are treated as the energy values assigned to each state (click state or non-click state). EBM defines the joint distribution of ($\x$, $y$, z) with the energy values: $\hat{p}(\x,y, z) = \frac{\exp(f_{\theta}(\x)[y])}{Z(\theta)}$, where $Z(\theta)$ is the unknown normalization constant. With such a definition, we prove that (in Section \ref{subsec::hdge}) $\ell_{\text{rank}}$ is approximately optimizes the contextualized generative objective
\begin{equation}
    \label{eq:rank-approx-generative}
    \ell_{\text{rank}} \approx -\sum_{\x, y, z} \log \hat{p}(\x|y, z),
\end{equation}
in which $\hat{p}(\x|y,z)$ stands for the probability of $\x$ occurring in the click samples (non-click samples) under the context $z$. Hence, JRC is approximately optimizing the contextualized hybrid discriminative and generative objective,
\begin{equation}
    \label{eq:context-loss-optimization}
    -\sum_{\x, y, z} \alpha \log \hat{p}(y|\x) + (1-\alpha) \log \hat{p}(\x|y,\z).
\end{equation}
The detailed analysis is offered in Section ~\ref{sec:link-with-ltr}.
In this sense, the new definition offers a unified interpretation for logit, and the click-through rate can be derived from the two logits.







Our later experiments validate the effectiveness of JRC in both offline and online experiments. By introducing the listwise-like generative loss, we see an improvement of ranking ability. Furthermore, consistent with previous studies on generative modeling~\cite{GrathwohlWJD0S2020JEM, Liu2020HDGE}, we also observe an improvement of calibration ability. One particularly interesting point is that, compared to the simple loss combination~\cite{sculley2010combined,li2015click}, our proposed approach further improves the model performance with the unified logit interpretation.


To summarize, the main contributions of our work are as follows:

\begin{itemize}
\item 
We propose the JRC approach for joint optimization of ranking and calibration in the task of click-through rate prediction. JRC extends the model output from the one-dimensional logit to two-dimensional logits, alleviating the conflict of multiple objectives. By optimizing the ranking of the click and non-click logits while constraining their subtraction to be the predicted probability, JRC effectively improves the ranking ability and keeps it well-calibrated.

\item We provide a deep analysis of the effectiveness of JRC and show that it unifies the interpretation of logits. By defining the joint distribution $\hat{p}(\x,y)$ with the new logits, we demonstrate that JRC approximately optimizes the contextualized hybrid discriminative and generative objective.
This unification provides a theoretical explanation for the effectiveness of JRC and also helps explain the efficacy of previous attempts on direct loss combination ~\cite{li2015click,sculley2010combined}. 

\item We verify the effectiveness of JRC in industrial datasets and examine the model performance through online A/B testing on the display advertising platform of Alibaba.
The experiment results show that JRC simultaneously improves the ranking and calibration abilities. 
During online A/B testing, JRC has also exhibited significant performance improvement, leading to a 4.4\% increase in CTR and a 2.4\% increase in RPM.
\end{itemize}

\section{Related Work}
We briefly review related studies from three aspects: click-through rate prediction, learning-to-rank, and hybrid modeling.

Most research efforts on \textbf{click-through rate prediction} have been devoted to improving model architectures. Wide \& Deep~\cite{cheng2016wide} and deepFM~\cite{guo2017deepfm} combine low-order and high-order features to improve model expressiveness. PNN~\cite{qu2016product} introduces a product layer to capture interactive patterns between inter-field categories. Considering the importance of historical user behaviors, DIN~\cite{zhou2018din} employs the attention mechanism to activate historical behaviors locally w.r.t. the target item and captures the diversity of user interest. 
Moreover, inspired by the success of the self-attention architecture in the tasks of sequence to sequence learning, Transformer is also introduced for feature aggregation~\cite{FengLSWSZY19DSIN}. 
Despite the recent advancement, less attention has been paid to defining a better loss function. Up to now, the pointwise model with a LogLoss has been the dominating paradigm for CTR prediction~\cite{zhou2018din}. The pointwise loss often yields well-calibrated predictions but may lead to a sub-optimal ranking performance~\cite{liu2011learning}. However, in practical industrial systems, both the calibration and ranking quality are essential when measuring the performance of a CTR prediction model~\cite{li2015click,sculley2010combined}. This requirement calls for research on direct optimization of the ranking.

A \textbf{Learning-To-Rank} (LTR) algorithm generally addresses the problem of ranking ability optimization. The main goal is to learn a scoring function for computing the degree of relevance, which further induces a ranking list~\cite{liu2011learning,XiaLWZL2008ListMLE,BruchWBN2019BinaryRelevance}. In its most straightforward format, a ranking problem can be viewed as correctly estimating the relevance score, and thus being named as the pointwise approach~\cite{gey1994inferring}. Another line of work has been focusing on correctly predicting ordered pairs by optimizing a pairwise loss~\cite{BurgesSRLDHH2005RankNet,GuSTRL2015RankingSVM}, which simplifies the learning-to-rank problem while misaligning with the ranking utilities. A ranking system is commonly evaluated with a set of ranking metrics such as Area Under the Curve~\cite{Fawcett2006ROC}, Normalized Discounted Cumulative Gain~\cite{jarvelin2002cumulated} or MAP (Mean Average Precision)~\cite{baeza1999modern}. Considering the gap between the optimization goal and evaluation metric, researchers have proposed a set of ranking-metric-based optimization approaches~\cite{CaoQLTL2007ListNet,qin2010general,burges2010ranknet}.

One critical issue for the non-pointwise LTR algorithms is that the prediction score no longer aligns with the click probability. To preserve both calibration and ranking quality, prior studies have examined a linear combination of pointwise and non-pointwise loss~\cite{li2015click,sculley2010combined,yan2022ScaleCalibration}. Despite being effective, the underlying meaning of the output is not well-understood as the two losses correspond to different goals. Recently, some studies have proposed to model the uncertainty of neural rankers and yield better uncertainty estimation~\cite{PenhaH2021CalibrationLTR,CohenMLRE2021CalibrationRetrieval,yan2022ScaleCalibration}. Unlike these works, this research focuses on the problem of jointly optimizing the ranking and calibration ability of CTR prediction models. To this end, we propose JRC and derive the pointwise and listwise-like losses. JRC offers a unified view of multiple losses, redefining logits as the energy of click/non-click states.

Under the above definition, we show that JRC is approximately modeling the contextualized \textbf{hybrid discriminative-generative}~\cite{RainaSNM2003ClassificationHybrid,Liu2020HDGE} objectives. In concrete, hybrid model combines generative and discriminative models to get the best of both paradigms. It has been observed that generative models benefit downstream problems such as calibration~\cite{GrathwohlWJD0S2020JEM,RainaSNM2003ClassificationHybrid}, semi-supervised learning~\cite{CSZ2006Semi,KingmaMRW2014SemiDeepGenerative} and the handling of missing data~\cite{NgJ2001OnDisGen}. By contrast, discriminative models are usually more accurate when labeled training data is abundant. By combining them, hybrid modeling achieves superior performance and produces more calibrated output. However, context information is largely ignored in previous studies. The proposed JRC method extends the idea of hybrid modeling with contextualization for CTR prediction. Incorporating context information further enables our model to be personalized and adaptable, which better meets online system requirements. 
In this work, we observe that the contextualized hybrid model JRC simultaneously improves the ranking quality and calibration, and the result is also consistent with prior studies~\cite{RainaSNM2003ClassificationHybrid,Liu2020HDGE,GrathwohlWJD0S2020JEM}.

\section{Methodology}

In the following, we give a brief description about the background firstly and then introduce the proposed JRC approach.

\subsection{Background}
In CTR prediction, the pointwise model takes the input as $(\x, y) \sim (X, Y) $, and learns a function $f$ with parameter $\theta$, $f_\theta: \mathbb{R}^D\rightarrow \mathbb{R}^1$, that maps $\x$ to a logit. Here, $\x$ denotes the feature and $y \in \{0, 1\}$ indicates the click label. The logit will then be used to compute the predicted click probability (pCTR for short) with the below sigmoid function:

\begin{align}
     \hat{p}(y=1|\x) = \frac{\exp(f_\theta (\x))}{1+\exp(f_\theta (\x))}.
\end{align}

In industrial applications like online advertising, calibrated predicted click probability is often required~\cite{li2015click}. Thus pointwise models are widely adopted as their predictions can be treated as the click probability~\cite{HastieFT2001ESL},  in which the LogLoss is minimized:
\begin{align}
     -\sum_{\x, y} \log \hat{p}(y|\x).
\end{align}

Pointwise models are well-calibrated but have inferior ranking ability. 
In this study, we propose a Joint optimization of Ranking
and Calibration approach (JRC). 

\subsection{Joint Optimization of Ranking and Calibration}
\label{sec:context-hybrid-method}
As mentioned above, previous attempts to improve ranking for CTR prediction model directly combines pointwise and pairwise losses ~\cite{sculley2010combined,li2015click}. Despite being effective, the meaning of the logit becomes unclear. To avoid representing different meanings with the same logit, the proposed JRC extends the output logit from 1 dimension to 2 dimensions, $f_\theta: \mathbb{R}^D\rightarrow \mathbb{R}^2,$ as shown in Figure~\ref{fig:hybrid}(c). The intuition of introducing an additional degree of freedom is to alleviate the conflict between the optimization of ranking and calibration.

Let $f_{\theta}(\x)[y]$ indicates the $y$-th index of $f_{\theta}(\x)$. In JRC, $f_{\theta}(\x)[1]$ is the logit corresponding the click state and $f_{\theta}(\x)[0]$ is the logit corresponds to the non-click state. JRC computes the predicted probability based on the subtraction $f_{\theta}(\x)[1] - f_{\theta}(\x)[0]$:
\begin{equation}
\begin{aligned}
  \hat{p}(y=1|\x) &=
     \frac{1}{1+\exp( -(f_{\theta}(\x)[1] - f_{\theta}(\x)[0]))}.
\end{aligned}
\end{equation}
Given the predicted probability, we firstly introduce the pointwise loss to preserve the calibration ability:
\begin{equation}
\begin{aligned}
  \ell_{\text{calib}} &= -\sum_{{\x,y}} \mathrm{log} ~\hat{p}(y|\x)\\ 
  &= -\sum_{{\x,y}} \log \frac{\exp(f_{\theta}(\x)[y])}
       {\exp(f_{\theta}(\x)[0]) + \exp(f_{\theta}(\x)[1])}.~\label{eq:calib-loss-2}
\end{aligned}
\end{equation}
Note that Equation~\ref{eq:calib-loss-2} is equivalent to the standard cross-entropy loss.
In recommendation, every item is displayed to the user in a specific context, e.g., the specific spot that item presented to users~\cite{ShengZZDDLYLZDZ2021STAR}.
To improve the relative ranking of the same context, we add a listwise-like loss aiming to learn the ranking in a specific context,
\begin{equation}
 \begin{aligned}
  \ell_{\text{rank}} = 
    -\sum_{\x, y, z} \log\frac{\exp(f_{\theta}(\x)[y])}{\sum_{\x_i\in X_{\z}}\exp (f_{\theta}(\x_i)[y])},
 \end{aligned}
 \label{eq:listwise-ranking-2}
\end{equation}
where $\z$ indicates the current context and $X_{\z}$ denote the set of samples that share the same context $\z$. 

The final objective of JRC can be written as:
\begin{equation}
 \begin{aligned}
  \ell_{\text{final}} = \alpha \ell_{\text{calib}} + (1-\alpha) \ell_{\text{rank}},
 \end{aligned}
\end{equation}
where $\alpha$ is the hyper-parameter between [0, 1]. In doing so, JRC optimizes the ranking ability through contrasting the logit value for a sample $x$ from the logit values of other samples with different labels, and constrains the predicted probability to be a monotonic function of the logit subtraction for calibration.

\subsection{Contextualized Hybrid Discriminative-Generative Model}
\label{subsec::hdge}
We further show that in JRC, the logits have a unified interpretation. In particular, JRC can be seen as an energy-based model (EBM)~\cite{lecun2006tutorial}.
In concrete, we treat the two-dimensional logits as energy values of $(\x, y, z)$ as  Equation~\ref{eq:joint_enegy_model}:
\begin{equation}
\hat{p}(\x,y,z) = \hat{p}(\x,y) = \frac{\exp(f_{\theta}(\x)[y])}{Z(\theta)},
\label{eq:joint_enegy_model}
\end{equation}
where  $Z(\theta)$ is the unknown normalization constant. The equation $\hat{p}(\x,y,\z)=\hat{p}(\x,y)$ holds since the context feature can be added to $\x$ thus we can absorb $\z$ into $\x$.
Given a specific context $\z$, $\hat{p}(\x,y|\z)$ is calculated as Equation~\ref{eq:context_enegy_model},

\begin{algorithm*}
\caption{A Tensorflow-style Pseudocode of our proposed JRC model.}
\label{alg:hybrid}
\begin{multicols}{2}
\begin{minted}[fontsize=\small]{python}
# B: batch size, label: [B, 2], context_index: [B, 1]
# Feed forward computation to get the 2-dimensional logits
# and compute LogLoss -log p(y|x, z)
logits = feed_forward(inputs)
ce_loss = mean(CrosssEntropyLoss(logits, label))

# Mask: shape [B, B], mask[i,j]=1 indicates the i-th sample
# and j-th sample are in the same context
mask = equal(context_index, transpose(context_index))

# Tile logits and label: [B, 2]->[B, B, 2]
logits = tile(expand_dims(logits, 1), [1, B, 1])
y = tile(expand_dims(label, 1), [1, B, 1])
# Set logits that are not in the same context to -inf
y = y * expand_dims(mask, 2)
logits = logits + (1-expand_dims(mask, 2))*-1e9
y_neg, y_pos = y[:,:,0], y[:,:,1]
l_neg, l_pos = logits[:,:,0], logits[:,:,1]

# Compute listwise generative loss -log p(x|y, z)
loss_pos = -sum(y_pos * log(softmax(l_pos, axis=0)), axis=0)
loss_neg = -sum(y_neg * log(softmax(l_neg, axis=0)), axis=0)
ge_loss = mean((loss_pos+loss_neg)/sum(mask, axis=0))

# The final JRC model
loss = alpha*ce_loss + (1-alpha)*ge_loss
\end{minted}
\end{multicols}
\end{algorithm*}

\begin{table*}[!t]
\centering
\renewcommand\arraystretch{1.5}
\caption{A comparison of listwise softmax loss and JRC generative loss with several toy examples.}
{
\begin{tabular}{c|c|c}
    \toprule 
    list (context)  & listwise softmax loss~\cite{CaoQLTL2007ListNet} & listwise generative loss (in JRC) \\
    \midrule
    condition a): ($d_i^+$, $d_j^+$)
    & $-(\log \frac{\exp{s_i}}{\exp{s_i}+\exp{s_j}} + \log \frac{\exp{s_j}}{\exp{s_i}+\exp{s_j}})$
    & $
    -(\log \frac{\exp{t_i^1}}{\exp{t_i^1}+\exp{t_j^1}}
      +
      \log\frac{\exp{t_j^1}}{\exp{t_i^1}+\exp{t_j^1}})
      $  \\
    \midrule
    condition b): ($d_i^-$, $d_j^-$)
    & 0
    & $
    -(\log \frac{\exp{t_i^0}}{\exp{t_i^0}+\exp{t_j^0}}
      +
      \log\frac{\exp{t_j^0}}{\exp{t_i^0}+\exp{t_j^0}})
      $  \\
    \midrule
    condition c): ($d_i^+$, $d_j^-$)
    & $-\log \frac{\exp{s_i}}{\exp{s_i}+\exp{s_j}}$
    & $
    -(\log \frac{\exp{t_i^1}}{\exp{t_i^1}+\exp{t_j^1}}
      +
      \log\frac{\exp{t_j^0}}{\exp{t_i^0}+\exp{t_j^0}})
      $ \\
    \bottomrule 
\end{tabular}
}
\label{tbl:link-with-ltr-table}
\end{table*}

\begin{equation}
\begin{aligned}
\hat{p}(\x,y|\z) &= \frac{\hat{p}(\x,y,\z)}{\hat{p}(\z)} = \frac{\hat{p}(\x,y)}{\hat{p}(\z)} \\
&= \frac{\exp(f_{\theta}(\x)[y])}{\sum_{\x_i\in \mathcal{X}_{\z}} \sum_{y^{\prime}}\exp(f_{\theta}(\x_i)[y^{\prime}])}.
\label{eq:context_enegy_model}
\end{aligned}
\end{equation}
 $\mathcal{X}_{\z}$ denote the full set of samples that includes $\z$.
Subsequently, we can compute the probability $\hat{p}(y|\z)$ by marginalizing over all $\x$ with context $\z$:
\begin{equation}
\hat{p}(y|z) = \frac{\sum_{\x_i\in \mathcal{X}_{\z}}\exp (f_{\theta}(\x_i)[y])}{\sum_{\x_i\in \mathcal{X}_{\z}} \sum_{y^{\prime}}\exp(f_{\theta}(\x_i)[y^{\prime}])}.
\label{eq:context_label}
\end{equation}
By marginalizing out $y$ in Equation~\ref{eq:joint_enegy_model}, we can also obtain the probability density $p(\x)$:
\begin{equation}
\hat{p}(\x) = \hat{p}(\x, z) = \frac{\sum_{y^{\prime}}\exp (f_{\theta}(\x)[y^{\prime}])}{Z(\theta)}.
\label{eq:nocontext_density}
\end{equation}
Then we can compute $\hat{p}(y|\x)$ via $\hat{p}(\x, y)/\hat{p}(\x)$ by dividing Equation~\ref{eq:joint_enegy_model} to Equation~\ref{eq:nocontext_density}:
\begin{equation}
\begin{aligned}
\hat{p}(y|\x) = &
  \frac{\hat{p}(\x, y)}{\hat{p}(\x)} =
  \frac{\exp(f_{\theta}(\x)[y])}
       {\sum_{y^{\prime}}\exp(f_{\theta}(\x)[y^{\prime}])}.\\
\label{eq:context_hybrid_discriminative}
\end{aligned}
\end{equation}
We can see it yields the standard softmax parameterization. 

Similarly, we can compute the $\hat{p}(\x|y, z)$ via $\hat{p}(\x, y|\z)/\hat{p}(y|\z)$ by dividing Equation~\ref{eq:context_enegy_model} to Equation~\ref{eq:context_label} as follows:
\begin{equation}
\begin{aligned}
\hat{p}(\x|y,\z) = &
  \frac{\hat{p}(\x, y|\z)}{\hat{p}(y|\z)} =
  \frac{\exp(f_{\theta}(\x)[y])}{\sum_{\x_i\in \mathcal{X}_{\z}}\exp (f_{\theta}(\x_i)[y])}. 
\label{eq:context_hybrid_generative}
\end{aligned}
\end{equation}

The denominator $\sum_{\x_i\in \mathcal{X}_{\z}}\exp (f_{\theta}(\x_i)[y])$ of Equation~\ref{eq:context_hybrid_generative} require summation over all sample with context $z$ that is intractable. Note that if we approximate the generative loss by replacing the full set of samples $\mathcal{X}_{\z}$ with samples in the mini-batch $X_{\z}$, then $-\log\hat{p}(\x|y,\z)$ has the same form as the listwise-like loss in JRC! Thus $\ell_{\text{rank}}$ is actually approximately optimizing the contextualized generative objective, i.e.,
$\ell_{\text{rank}} \approx -\sum_{\x, y, z} \log p(\x|y, z).$
In the later experiment section, we compare the different context choices and give a detailed analysis.


To sum up, we show that JRC has a novel objective function for CTR prediction that consists of both discriminative loss and context-dependent generative loss. 
Unlike non-pointwise LTR approaches, JRC first defines the joint probability and then derives the conditional probabilities, preserving the meaning of output $\hat{p}(y|\x)$ as click-through rate for calibration. 
In the later experiment section, we show that the proposed JRC gains the benefits of both discriminative and generative approaches, improving both ranking and calibration abilities.
The TensorFlow-like~\cite{tensorflow2015-whitepaper} pseudo code  of JRC  is presented in Algorithm~\ref{alg:hybrid}.

\subsection{Understanding JRC}
\label{sec:link-with-ltr}
For the discriminative loss in Equation~\ref{eq:context_hybrid_discriminative}, the subtraction between the two logits of JRC is equivalent to the one-dimensional logit in the pointwise approach.
Interestingly, we observe that the listwise generative loss in Equation~\ref{eq:context_hybrid_generative} is structurally similar to the listwise softmax loss. They share the same goal: contrasting the logit value for $\x$ with label $y$ from the logit values of other training samples in the same context list. 

To better illustrate the connections and distinctions between the listwise softmax loss and the generative loss of JRC, we provide several toy examples in Table~\ref{tbl:link-with-ltr-table}. For each candidate $d_i$, we represent the corresponding listwise softmax logit as $s_i$, and the JRC logits as ($t_i^0$, $t_i^1$). Here, $t_i^0$ (and $t_i^1$) refers to the logit for the non-click (and click) state. We further represent the item with a positive (click) or a negative (non-click) label as $d^+$ and $d^-$, respectively. We take into account all of the three possible conditions for a list: a) all items are clicked, b) all items are non-clicked, and c) partial items are clicked. The list (for the listwise model) and the context (for our JRC model) are aligned for better comparison.

Table~\ref{tbl:link-with-ltr-table} shows that in list a), the listwise softmax loss is structurally similar to the listwise generative loss. Meanwhile, in list c), the listwise loss is also an essential component of the generative loss. However, they differ substantially in list b) as JRC explicitly models the likelihood of generating the non-click data, which is overlooked in listwise models. One potential benefit of this approach is that it helps smooth the predictions of non-clicked samples, providing regularization for such data, thus improving the calibration. In addition, this also makes it possible to utilize non-click information that is usually abundant for industrial systems. It is worth noting that despite similar, the underlying definition for logit remains to be different between JRC and listwise models.

\section{Experiment}
In this section, we conduct extensive experiments to understand the effectiveness of the JRC model.

    

\subsection{Experiment Setup}
\subsubsection{Datasets}
Our experiments are conducted on two widely-adopted public datasets~\cite{zhu2021open} and one production dataset. 
The basic statistics of these datasets are summarized in Table~\ref{tab:dataset}.

\begin{itemize}
    \item 
    \textbf{Avazu}\footnote{https://www.kaggle.com/c/avazu-ctr-prediction/data}. 
    Avazu is a Kaggle challenge dataset for CTR prediction. The data comes from a mobile advertising platform named Avazu and includes 22 features, such as device information, advertisement category, and other attributes. It consists of ten days data for training and validation and one day of data for testing. 
    We use the benchmark data version provided by Zhu et al.~\cite{zhu2021open}.
    
    \item 
    \textbf{Taobao}\footnote{https://tianchi.aliyun.com/dataset/dataDetail?dataId=649}. 
    This dataset provides various user behavior information, including click, purchase, add-to-cart, and favorite, for about one million users from the Taobao recommendation system. It contains eight days of click-through data, with the first seven days for training and validation and the rest one day for testing. A few pre-processing steps are carried out by following~\cite{BianWRPZXSZCMLX2022CAN}.

    \item
    \textbf{Production}. The production dataset is sampled from the impression log of Alibaba online system. The full impression and click data from 2021/11/24 is adopted for training, and a subset of data from the next day is used for testing. The training set consists of 1.8 billion samples with 286 features, and the testing set contains 30 million samples.
\end{itemize}

{
\begin{table}[!tbp]
\centering
\caption{Statistics of the public and production datasets.}
\label{tab:dataset}
\resizebox{\columnwidth}{!}
{
\begin{tabular}{c|ccc}
	\toprule
	Dataset& \#Training samples& \#Testing samples & \#Features \\
	\midrule
	Avazu       &   28M     & 8M    & 22 \\
	Taobao      &   0.7M    & 0.3M  & 5\\
	Production  &   1.8B    & 30M   & 286\\
	\bottomrule
\end{tabular}
}
\end{table}
}

\subsubsection{Baselines.}
We include five baseline methods for comparison. According to the adopted loss function, we divide them into three groups. The first group (\textbf{G1}) is the pointwise model optimizing the LogLoss for each sample, and as a matter of fact, this is also the one deployed in the production system. The second group (\textbf{G2}) involves a pairwise model and a listwise model that directly optimize the ranking loss. Here, we utilize RankNet~\cite{burges2010ranknet} for pairwise loss and ListNet~\cite{CaoQLTL2007ListNet} for listwise loss. They are adopted due to their superior performance and simplicity for implementation~\cite{bruch2020stochastic,pasumarthi2019tf}. The third group (\textbf{G3}) considers both ranking loss and calibrated pointwise loss, but fuses them with a linear combination~\cite{li2015click,sculley2010combined}. 

\begin{itemize}
    \item[--] \textbf{G1: Pointwise} model to optimize the standard LogLoss~\cite{liu2011learning}. This is also our production baseline.
    \item[--] \textbf{G2: RankNet}~\cite{BurgesSRLDHH2005RankNet} that employs a pairwise loss to maximize the probability of correct ranking for pairs of samples.
    \item[--] \textbf{G2: ListNet}~\cite{CaoQLTL2007ListNet} which defines a listwise loss to maximize the likelihood of generating the correct ranking list. 
    \item[--] \textbf{G3}: The \textbf{Combined-Pair} approach that derives a new loss by a linear combination of pointwise and pairwise loss~\cite{li2015click}.
     \item[--] \textbf{G3}: The \textbf{Combined-List} approach that combines the pointwise and the ListNet~\cite{CaoQLTL2007ListNet}.
\end{itemize}

\begin{table*}[!t]
\centering
\caption{A comparison of model performance on two widely-adopted public datasets. The best results are highlighted in bold. A number with * indicates that the improvement over the \textit{Combined-Pair} and \textit{Combined-List} methods is significant with p-value < 0.05 by pairwise t-tests.}
\label{tab:publicperformance}
		\begin{tabular}{c|c|cccc|cccc}
			\toprule
		  \multicolumn{2}{c|}{\multirow{2}{*}{Dataset}} &  \multicolumn{4}{c|}{Avazu} & 
		  \multicolumn{4}{c}{Taobao} \\
		  \cline{3-10}
		  \multicolumn{2}{c|}{} & {AUC} & {LogLoss} & {PCOC} & {ECE} & {AUC} & {LogLoss} & {PCOC} & {ECE}\\
		\midrule		 
		 {\textbf{G1}} & Pointwise & 0.7620 & 0.3683  & \textbf{1.0163} & \textbf{0.0042}  & 0.8723 & 0.2223 & 1.019 & 0.0217\\
		 \midrule
		 \multirow{2}{*}{\textbf{G2}} & RankNet~\cite{burges2010ranknet} &0.7639
  &0.5939 & 2.8256 & 0.2784 &0.8749  &0.2210  &0.9632  &0.0233 \\
		 & ListNet~\cite{CaoQLTL2007ListNet} &0.7646  &1.1393 &4.4817  & 0.5309
 &0.8733  &0.4931  &1.7427  &0.3899 \\
		  \midrule
		  \multirow{2}{*}{\textbf{G3}} 
		  &  Combined-Pair~\cite{li2015click} &0.7644&
0.3670
  &1.0330
 & 0.0068
 & 0.8740  & 0.2209 &
1.025  & 0.0145\\
& Combined-List & 0.7643 & 0.3694 & 1.0513 & 0.0078 & 0.8729 & 0.2232 & 0.9947 & 0.0178 
  \\
		  \midrule
		  {\textbf{Ours}} & JRC &\textbf{0.7649*}  &\textbf{0.3667} &1.0276*  & 0.0065* &\textbf{0.8765*}
& \textbf{0.2189*} & 
\textbf{0.9953} & 
\textbf{0.0120*}\\
		 \bottomrule
		\end{tabular}
\end{table*}

\subsubsection{Evaluation Metrics.}
For evaluating the ranking performance, we adopt the standard \textbf{AUC} metric (Area Under Receiver Operating Characteristic Curve). A larger AUC indicates a better ranking ability. Note that in the production dataset, we further compute the Group AUC (\textbf{GAUC}) to measure the goodness of intra-user ranking ability, which has shown to be more consistent with the online performance~\cite{ZhuJTPZLG2017OCPC,zhou2018din,ShengZZDDLYLZDZ2021STAR}. GAUC is also the top-line metric in the production system. It can be calculated with Equation~\ref{eqn:gauc}, in which $U$ represents the number of users, $\#\textrm{impression}(u)$ denotes the number of impressions for the $u$-th user, and $\textrm{AUC}_u$ is the AUC computed only using the samples from the $u$-th user. However, due to the lack of user information in the Avazu dataset, and limited samples per user in Taobao, we cannot effectively compute the GAUC metric; therefore, the AUC metric is reported for the public datasets instead.
\begin{equation}
\label{eqn:gauc}
\begin{aligned}
    \textrm{GAUC} = \frac{\sum_{u=1}^U \# \textrm{impressions}(u) \times \textrm{AUC}_u}{\sum_{u=1}^U \# \textrm{impressions}(u)}
\end{aligned}
\end{equation}

To measure the calibration performance of each method, we exploit a set of metrics including the averaged \textbf{LogLoss}, the expected calibration error (\textbf{ECE}), and the predicted CTR over the true CTR (\textbf{PCOC})~\cite{GuoPSW2017OnCalibration,PanATLLXH2020ProbECE}. They are all defined in Equation~\ref{eqn:eval-metric}. For the $i$-th sample, let $\hat{p}_i$ represent the pCTR. 
For ECE, we first partition the range [0, 1) equally into $K$ buckets. $\vmathbb{1}(\hat{p}_i\in{B_k})$ is an indicator with a value of 1 if the predicted probability locates in the $k$-th bucket $B_k$, and otherwise 0. LogLoss measures the sample-level calibration error, whereas PCOC and ECE provide a dataset-level and a subset-level calibration measurement. A lower LogLoss or ECE implies a better performance; for PCOC, our goal is to obtain a value close to 1.0.
{
\begin{equation}
\label{eqn:eval-metric}
\begin{aligned}
    \text{LogLoss} &= -\frac{1}{\mathcal{D}}\sum_{i=1}^{\mathcal{D}} \big(y_i\log\hat{p}_i + (1 - y_i)\log(1-\hat{p}_i)\big) \\
    \text{ECE} &= \frac{1}{\mathcal{D}}\sum_{k=1}^{K}|\sum_{i=1}^{\mathcal{D}}(y_i-\hat{p}_i)\:\vmathbb{1}(\hat{p}_i\in{B_k})|\\
    \text{PCOC} &= \frac{1}{\sum_{i=1}^{\mathcal{D}} y_i}\sum_{i=1}^{\mathcal{D}} \hat{p}_i
\end{aligned}
\end{equation}
}
\subsubsection{Implementation Details.}
In both Taobao and the production datasets, we utilize \textbf{CAN}~\cite{BianWRPZXSZCMLX2022CAN} as the neural network structure for all methods. CAN has been shown to be effective for CTR prediction with user behavior sequence. However, due to a lack of user behavior sequence information, for the Avazu dataset, we instead employ the \textbf{DeepFM}~\cite{guo2017deepfm} structure, which has also shown to achieve an on-par performance as CAN~\cite{zhu2021open}. For all datasets, each method is trained with one epoch, following~\cite{ZhangSZJHDZ2022OneEpoch}. For the production dataset, all of our experimental models are trained with the XDL platform~\cite{zhang2022picasso}.

For JRC, we adopt the Adam optimizer with the initial learning rate of 0.001 for Avazu and 0.004 for Taobao. For hyper-parameters, we experiment with the weight ratio $(1-\alpha) / \alpha$ rather than a direct tuning of $\alpha$. The weight ratio is set to 0.01 for Avazu and 1.0 for Taobao. Again, due to the lack of user information and the sparsity of the same-user data samples in Taobao and Avazu, we cannot offer a meaningful, fine-grained definition for the context $X_z$. 
Therefore, we set $X_z$ as sample of the same min-batch for the public datasets. 
Whereas for the production dataset, we will experiment with the influence of different definitions of the context.

\subsection{Performance on Public Datasets}
\subsubsection{Overall Performance}
Table~\ref{tab:publicperformance} provides the model performance for all the compared methods. Overall, we observe an improved ranking ability by adopting the pairwise or listwise model in both datasets. However, this causes a significant deterioration of calibration-related metrics, mainly ascribed to the fact that ranking loss-based optimization only focuses on the relative order of samples and pays no attention to the absolute value of the prediction. The prediction no longer corresponds to the click probability, thus yielding unsatisfying calibration results. The Combined-Pair and Combined-List approaches provide a rescue that brings back the calibration ability while maintaining the ranking performance at the same level. Note that this result is also consistent with prior studies~\cite{li2015click,sculley2010combined,pasumarthi2019tf,bruch2020stochastic}.

For the JRC model, we observe a further improvement of ranking and calibration performance over the Combined-Pair and Combined-List methods, and most of the metric improvement is statistically significant. The result validates the effectiveness of our unified hybrid framework. Compared with the combined methods, JRC offers a consolidated view of the pointwise loss and listwise-like generative loss. It also effectively utilize the huge amount of non-click data. By contrast, the combined method imposes irrelevant assumptions on the same logit, limiting the optimization and leading to sub-optimal solutions.


\subsubsection{Impact of hyper-parameter}
\begin{figure*}[!tbp]
\centering
\subfigure[]{
\includegraphics[width=.32\linewidth]{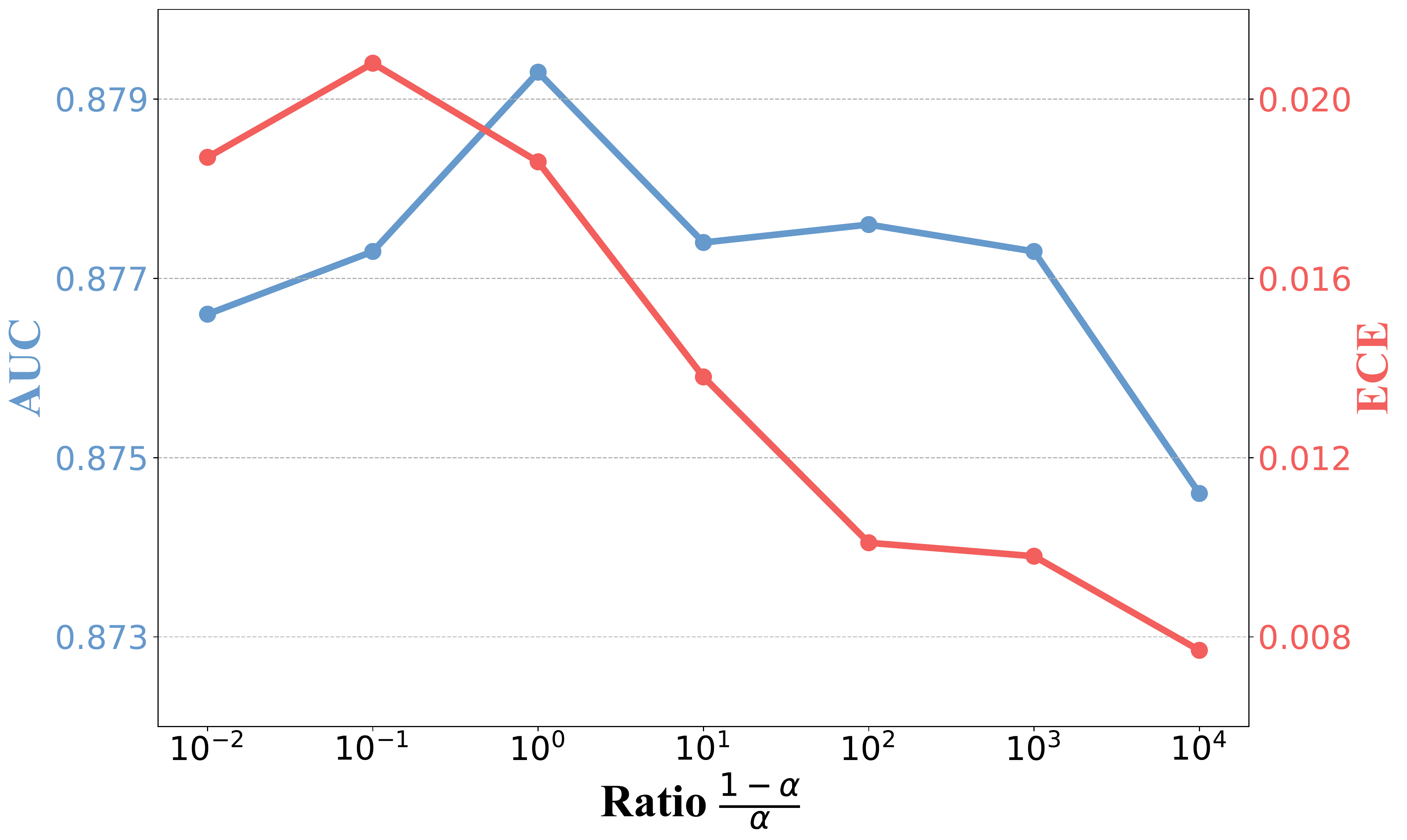}\label{fig:weight_ratio}
}
\subfigure[]{
\includegraphics[width=.32\linewidth]{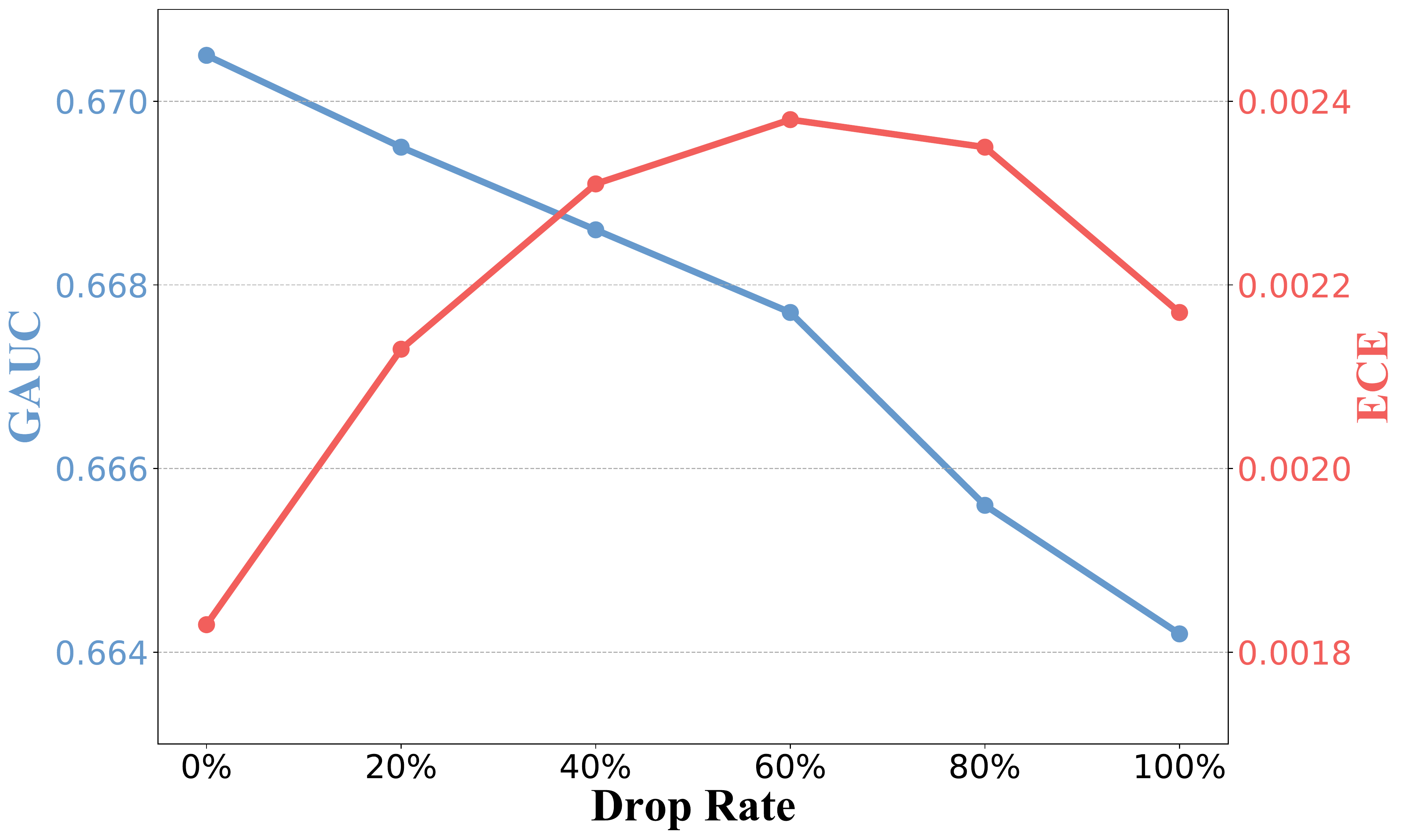}\label{fig:eval_group_size}
}
\subfigure[]{
\includegraphics[width=0.32\linewidth]{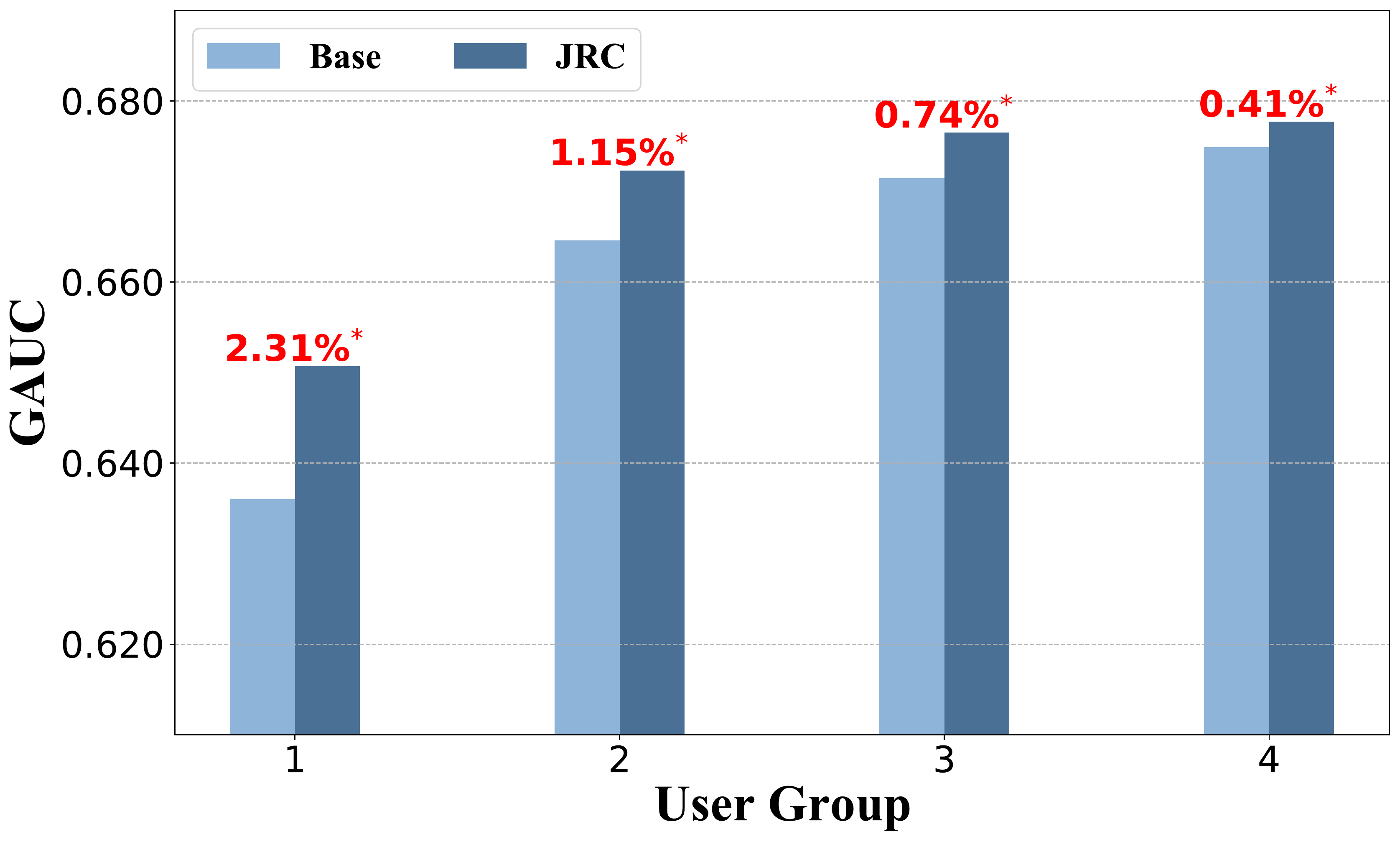}\label{fig:eval_user_group}
}
\caption{(a) Performance of JRC w.r.t. different ratios $\frac{1-\alpha}{\alpha}$ on the Taobao dataset. (b) The influence of context size for the proposed JRC method. 
(c) Performances of Baseline (pointwise) and JRC model on different user groups. Here, we compute the relative performance improvement over the baseline, $*$ denotes the improvement is statistically significant. 
}
\end{figure*}

We investigate how the ranking and calibration abilities change with various weighting ratios between the discriminative loss and the listwise generative loss of JRC. The result is illustrated in Figure~\ref{fig:weight_ratio}. For simplicity, we only experiment with the Taobao dataset. To be specific, we first re-scale the two loss values to the same level and then vary $\frac{1-\alpha}{\alpha}$ in [$10^{-2}$,$10^{-1}$,...,$10^{4}$] with all of the other hyper-parameters fixed. We find that an extremely large or small value of $\frac{1-\alpha}{\alpha}$ would hurt the ranking ability as one loss largely dominates over the other, and JRC degenerates to a pointwise model or a listwise-like model. A weighting ratio in between help achieve the best performance, which again validates the importance of hybrid modeling. We also observe that ECE generally decreases with the increase of $\frac{1-\alpha}{\alpha}$, indicating that a more emphasis on the generative loss can yield improved calibration performance.


\subsection{Performance on Production Dataset}
\begin{table}[!tbp]
\renewcommand\arraystretch{1.2}
\centering
\caption{A comparison of model performance for different methods on the production dataset.  The best results are highlighted in bold.}\label{tab:performance_production}
		\resizebox{\columnwidth}{!}{
		\begin{tabular}{c|c|cccc}
			\toprule
		  \multicolumn{2}{c|}{\multirow{2}{*}{Dataset}} &  \multicolumn{4}{c}{Production} \\
		  \cline{3-6}
		  \multicolumn{2}{c|}{} & {GAUC} & {LogLoss} & {PCOC} & {ECE}\\
		  \midrule		 
		 {\textbf{G1}} & Pointwise & 0.6642 & 0.3005 & 1.1156 & 0.0021\\
		 \midrule
		 \multirow{2}{*}{\textbf{G2}} & 
		 RankNet (session)~\cite{burges2010ranknet}& 0.6710 & 2.8118 & 25.016 & 0.4517 \\
		 & ListNet (session)~\cite{CaoQLTL2007ListNet} & \textbf{0.6714} & 3.1603 & 27.163 & 0.4884 \\
		  \midrule
		  \multirow{1}{*}{\textbf{G3}} 
		  & Combined-Pair (session)~\cite{li2015click} & 0.6673 & 0.3004 & 1.1279 & 0.0024 \\
		 \midrule
		  \multirow{1}{*}{\textbf{Ours}} 
		      & JRC (session) &0.6705 & \textbf{0.3004} & \textbf{1.0975} & \textbf{0.0018} \\
		 \bottomrule
		\end{tabular}
	}
\end{table}

\subsubsection{Overall Performance}
In the production dataset, we compare JRC with the pointwise model (the one deployed in the production system), RankNet, ListNet, and one representative combined method - Combined-Pair. Instead of AUC, we use Group AUC (GAUC) to measure the goodness of intra-user ranking quality, which is more relevant to online performance in our production system.
Table~\ref{tab:performance_production} offers the comparison of model performance. Note that for RankNet, ListNet, and Combined-Pair, we choose samples in the same \textit{session} to construct the pairs/list. We define a \textit{session} as all of the ads a user receives within a specific time window (e.g., ten minutes).

Like the public datasets, RankNet and ListNet outperform the Pointwise model for ranking ability, but at the expense of hurting well-calibrated probabilities. The negative impact of RankNet and ListNet on calibration makes them improper for CTR prediction.
The Combined-Pair method improves the ranking ability while maintaining the comparable calibration performance to the pointwise model. The result again demonstrates the effectiveness of combining the pointwise and pairwise losses. JRC (session) improves ranking and calibration metrics over the Combined-Pair method. A detailed analysis of the context definition for JRC is provided in the following.

\begin{table}[!tbp]
\centering
\caption{A comparison of different context for the JRC model on the production dataset. The best results are highlighted in bold.}\label{tab:performance_jrc_context}
		\begin{tabular}{c|cccc}
			\toprule
		 Method  & {GAUC} & {LogLoss} & {PCOC} & {ECE}\\
		  \midrule	
		     Pointwise & 0.6642 & 0.3005 & 1.1156 & 0.0021 \\
		     \midrule	
             JRC (batch) & 0.6654  & 0.3006 & 1.1575 & 0.0029 \\
             JRC (gender) & 0.6651  & 0.3005 & 1.1293 & 0.0024 \\
		     JRC (domain) & 0.6661  & 0.3005 & 1.1344 & 0.0025\\
		 \midrule
		      JRC (session) &\textbf{0.6705} & \textbf{0.3004} & \textbf{1.0975} & \textbf{0.0018} \\
		 \bottomrule
		\end{tabular}
\end{table}

\subsubsection{Defining Context.}\label{sec:define-context}
We show that the definition of contexts, i.e., the context list to compute the listwise generative loss,  is essential to the final performance of JRC.
We consider four different context definitions -- batch, gender (of the user),  domain (i.e., different spots an ad can be placed on), and session for the JRC method. 
We first experiment with the batch context, in which samples from the same mini-batch are treated as being in the same context. In terms of other contexts, we treat data samples from the same mini-batch and within the same user gender/domain/session as the same context.

Table~\ref{tab:performance_jrc_context} offers a detailed comparison.
JRC (batch) does not improve much compared with the pointwise approach, demonstrating that a too broad context might introduce irrelevant samples. By defining the context as user gender, we do not see much performance improvement, implying that users of the same gender may not share similar interests. Besides, JRC (domain) achieves better performance than JRC (batch), suggesting the application domain is a meaningful context for consideration. A finer-grained context definition with JRC (session) achieves the best performance, illustrating that samples within the same user are more proper to represent a user's current interest and used as the context. 
\subsubsection{Influence of the Context Size.}\label{sec:context-size} 
We also study the effect of context size, i.e., number of samples per context, on the listwise generative loss. Here we conduct experiments on JRC (session). The average number of samples in each session is 6.8 on the production dataset. For each session, we randomly drop $\{20\%, 40\%, 60\%, 80\%, 100\%\}$ of samples, and compute the generative loss to train the model. Note that the random drop only affects the listwise generative loss, whereas the pointwise loss keeps intact. 

The result is shown in Figure~\ref{fig:eval_group_size}. We observe that the best ranking and calibration metrics are obtained when the drop rate is 0\%. Meanwhile, the ranking ability gradually decreases when we raise the drop ratio, suggesting that larger context size help improve the ranking ability. The finding is consistent with previous work~\cite{li2015click} that more effective pairs help induce a better ranker for the pairwise loss. 

A particularly interesting observation is that the calibration ability gradually decreases and then rises with the increase of the drop rate. We hypothesis that this is because when the drop rate is small, the noise during the computation of generative loss determines the calibration ability. However, when the drop ratio increases further, the generative loss becomes smaller compared with the discriminative loss, which means the discriminative loss becomes dominant. The dominant discriminative loss makes the calibration better and becomes similar to the pointwise model gradually.
\subsubsection{Performance on Different User Groups}
To further understand the behavior of the proposed JRC model, we split users into four groups based on their activity levels (which is measured by the numbers of clicks in the past 14 days). Users in Group 1 have the least historical behaviors, and users in Group 4 have the most behaviors. It is worth noting that users are grouped in this way to ensure that different user groups contain the same amount of data samples. Afterward, we compute the relative GAUC improvement (GAUC$_{JRC}$-GAUC$_{Base}$)/GAUC$_{Base}$ over the pointwise model for each group. 


As shown in Figure~\ref{fig:eval_user_group}, JRC consistently outperforms the pointwise model over all four user groups, demonstrating the effectiveness of our model over different types of users. Meanwhile, we see more improvement for users with few behaviors, whereas there is less improvement for users with more behaviors. This result is consistent with previous studies on generative modeling, which often yields better performance when the training data is limited ~\cite{RainaSNM2003ClassificationHybrid,NgJ2001OnDisGen}. In our case, samples for inactive users are more sparse. The generative part of JRC adds more training signals for users with few click behaviors, as it includes an objective that contrasts the positive samples with negative samples. Note that we can also connect this finding to a previous similar study conducted in Twitter Timeline~\cite{li2015click} observing that the combination of pointwise loss and ranking loss helps alleviate the sparsity of training signal in CTR prediction.


\subsection{Online Deployment}
\subsubsection{Online Training.} 

\begin{figure}[!t]
	\centering
	\includegraphics[width=\columnwidth]{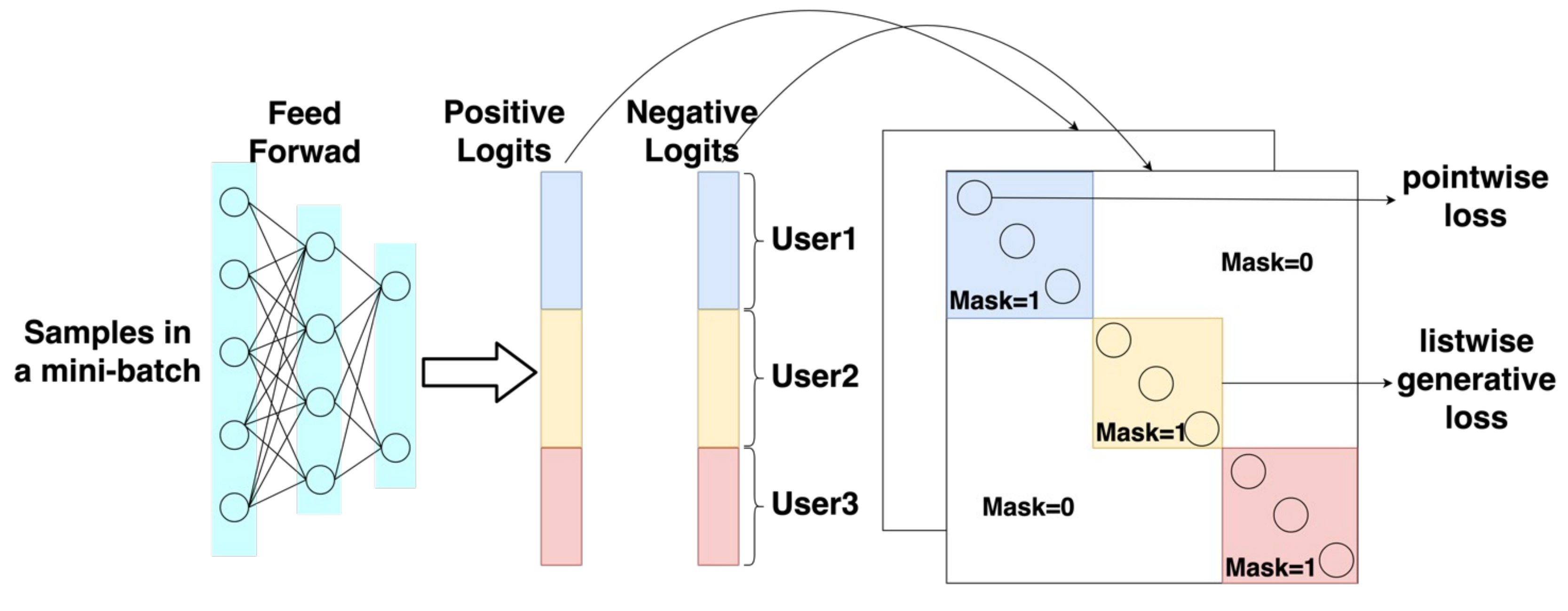}
	\caption{Samples from the same user are grouped  within the time window of 10 minutes and fed to the model to calculate the logits. These logits are used to compute the pointwise loss and listwise generative loss.}
	\label{fig:online-training}
\end{figure}

One major challenge for industrial recommendation systems is dealing with the constant shifting of data distribution. To keep up with data change, industrial systems need to update the model  continuously~\cite{KtenaTTMDHYS2019OnlineDFM,GuSFZZ2021Defer,chan2023capturing}. Due to the large amount of data, industrial CTR models are usually trained in a distributed manner. In this manner, samples of the same context might be sent to different workers as the training data comes as a stream. Note that the modification of the data organization will not affect the pointwise model but may greatly influence the proposed JRC method.

To support the \textbf{context-aware online training}, we redesign the training data organization so that samples of the same context are grouped into one mini-batch. Specifically, we first set a time window, and samples in the same context and occurring in the same time window will be placed into one mini-batch. Note that there is a trade-off -- an extended time window increases the context size, while the data delay is severe, and the resources to cache samples also increase dramatically. In practice, \textbf{samples of the same user are grouped within the time window of 10 minutes}, the illustration of the loss computation during online training is shown in Figure~\ref{fig:online-training}.

In addition, to serve the JRC model training, the context-sensitive organization of training samples also helps reduce data redundancy. Features of the same users will only be stored once and not repeatedly multiple times in different samples. Such an organization can compress our training data to a great extent, largely relieving the burden of system storage and computation.

\subsubsection{Online Performance.} 
The proposed JRC model is also deployed online for A/B testing in the display advertising system of Alibaba.
Our experiment buckets are determined by randomly hashing the unique identifier of a user.
The production baseline is a pointwise model. Compared with the pointwise model, JRC utilizes the same set of features and MLP structures, thus adding no additional computation cost for online serving.
We illustrate the performance lift of JRC in Table~\ref{tab:online}. Through a five-day of online experiments (from 2022/05/21 - 2022/05/25), 
JRC obtains a significant performance gain over the production baseline, with a +4.4\% increase in CTR, a +2.4\% increase in RPM (Revenue Per Mille), and a -0.27\% drop of LogLoss.

\begin{table}[!tbp]
\centering
\large
\caption{The performance lift of the proposed JRC model compared with the production baseline.}
\label{tab:online}
{
\begin{tabular}{c|ccc}
	\toprule
	 Metric       &   CTR     &   RPM     & LogLoss \\
	\midrule
	Lift    &  +4.4\%  &  +2.4\%  & -0.27\%\\
	\bottomrule
\end{tabular}
}
\end{table}

\section{Conclusion and Discussion}
Ranking ability and calibration ability are two essential aspects of CTR prediction. Pointwise models yield calibrated outputs but fall short of ranking ability. The pairwise and listwise models lift the ranking performance but hurt the calibration ability. A direct combination of the two models balances ranking and calibration. However, it breaks the meaning of output as the click probability, leading to suboptimal solutions. 

In contrast to the above models,  we propose the JRC method, which employs two logits corresponding to click and non-click states and jointly optimizes the ranking and calibration abilities. 
We further showed that JRC unified the logit interpretation as the energy-based model of the joint distribution. On top of that, conditional probabilities for discriminative and listwise generative losses can be derived.  
Our experiments on both public datasets, production datasets, and online A/B testing proved the effectiveness of JRC. Moreover, the proposed model further improves the ranking and calibration performance compared to a linear combination of multiple losses.


\bibliographystyle{ACM-Reference-Format}
\bibliography{sample-base}


\end{document}